\documentclass[aps,prl,superscriptaddress,10pt,twocolumn,amsmath,floatfix]{revtex4}

\usepackage{graphicx}
\usepackage{amsmath, amsthm, amssymb}

\begin{document}
\title{A pattern-forming instability co-driven by distinct mechanisms increases pattern diversity}
\author{Shai Kinast}
\author{Yuval R. Zelnik}
\author{Golan Bel}
\affiliation{Department of Solar Energy and Environmental Physics, Blaustein Institutes for Desert Research, Ben-Gurion University of the Negev, Sede Boqer Campus 84990, Israel}
\author{Ehud Meron}
\affiliation{Department of Solar Energy and Environmental Physics, Blaustein Institutes for Desert Research, Ben-Gurion University of the Negev, Sede Boqer Campus 84990, Israel}
\affiliation{Department of Physics, Ben-Gurion University, Beer
Sheva, 84105, Israel}
\date{\today}

\begin{abstract}
We use the context of dryland vegetation to study a general problem of complex pattern forming systems - multiple pattern-forming instabilities that are driven by distinct mechanisms but share the same spectral properties. We find that the co-occurrence of such instabilities results in the growth of a single mode rather than two interacting modes. The interplay between the two mechanisms, which promote or counteract each other, compensates for the simpler dynamics of a single mode by inducing higher pattern diversity. Possible implications to biodiversity of ecosystems are discussed.
\end{abstract}

\maketitle

Instabilities of uniform states in complex pattern-forming systems can be driven by two or more independent physical mechanisms. An illuminating example is vegetation pattern formation in water-limited systems (drylands).
There is an increasing evidence that dryland landscapes can self organize to form spatial vegetation patterns even in fairly uniform regions~\cite{Valentin1999catena,Deblauwe2008geb}. Vegetation pattern formation is driven by positive feedbacks between local vegetation growth and water transport towards the growing vegetation. The depletion of water in the vicinity of the growing vegetation inhibits the growth there and promote nonuniform vegetation growth~\cite{Meron2012eco_mod}.
At least three mechanisms of water transport can be distinguished. Overland water flow induced by higher infiltration rates in denser vegetation patches,
water conduction by laterally extended root zones that further extend as the plants grow, and fast soil-water diffusion, relative to biomass expansion, in conjunction with strong water uptake by confined root zones.
In the following we refer to the positive feedbacks associated with these transport mechanisms as to the ``infiltration'', ``root-augmentation'' and ``uptake-diffusion'' feedbacks, respectively.

\begin{figure*}[!ht]
  \centering
  \includegraphics[width=0.9\linewidth]{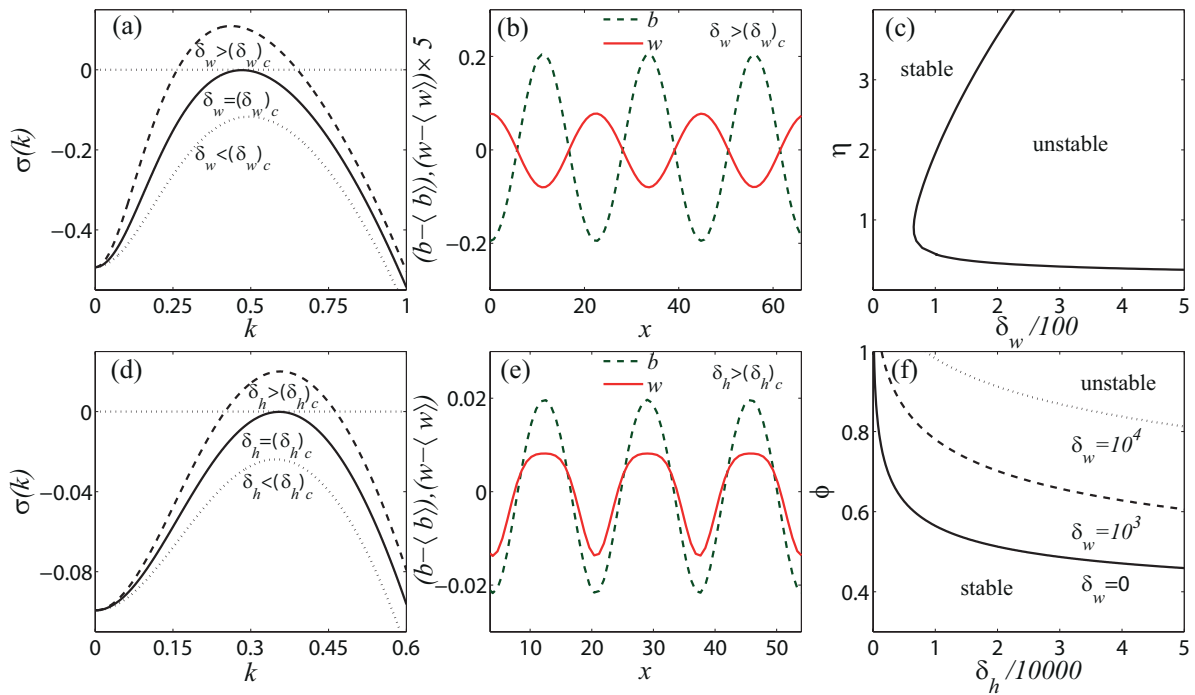}
  \caption{
  Nonuniform stationary instabilities of uniform vegetation driven by the uptake-diffusion feedback ($\phi=0$) (a,b,c) and by the infiltration feedback ($\eta=0$) (d,e,f). Panels (a) and (d) show the growth rates, $\sigma(k)$, of periodic perturbations with wavenumbers $k$ below (dotted line) at (solid line) and beyond (dashed line) the instabilities. Panels (b) and (e) show the anti-phase and in-phase patterns the instabilities lead to, and panels (c) and (f) show the instability thresholds in the planes spanned by the parameters that control the instabilities. The three lines in panel (f) correspond to different values of $\delta_w$ as indicated in the figure; although $\eta=0$ the instability threshold depends on the soil-water diffusivity. In panel (c) there is only one curve because for $\phi=0$ the instability threshold is independent of the water transport coefficient, $\delta_h$.  
  Parameters for panels (a,b): $\phi=0$, $\eta=0.9$ and $\delta_w=70$.
  Parameters for panels (d,e): $\eta=0$, $\delta_w=0$, $\phi=0.9$ and $\delta_h=400$.}
  \label{fig:profiles2}
\end{figure*}

The instabilities induced by the different feedbacks all share the \emph{same} spectral properties, that is, they all lead to monotonously growing modes that have the same spatial symmetry, a finite-wavenumber mode in 1d (Fig. \ref{fig:profiles2}), or in 2d, the simultaneous growth of three modes with wave-vector directions $2\pi/3$ apart that yield hexagonal patterns.
However, the modes that grow at these instabilities, and consequently the patterns that emerge, differ in the relative biomass-water distributions. The infiltration feedback acts to increase the soil-water content in patches of denser biomass and therefore leads to in-phase biomass-water patterns~\cite{Rietkerk2002an}. By contrast, the root-augmentation feedback and the uptake-diffusion feedbacks act to deplete the soil water content in denser biomass patches, because of the higher water uptake, and therefore lead to anti-phase biomass-water patterns~\cite{Gilad2007jtb}.

Although the three feedbacks represent independent mechanisms of vegetation pattern formation they are related to one another in the sense that varying the strength of one feedback may affect the strength of a different feedback. As a consequence, codimension-2 points~\cite{fn1} can be identified where two instabilities induced by distinct mechanisms coincide. The interplay between two co-occurring instabilities has been studied extensively for cases where the growing modes differ in their spectral properties (as dictated by the eigenvalues of the linear problem), i.e. either in their growth form, monotonic or oscillatory, or in their spatial symmetry or in both. Such co-occurring instabilities are known as ``codimension-2 bifurcations''~\cite{fn2}.
An illustrative example is the Hopf-Turing bifurcation in which a spatially periodic mode grows monotonically in time along with a uniform mode that grows in an oscillatory manner~\cite{Wit1996pre,Tlidi1997pre}.
Another example is the growth of two surface-wave modes that have different spatial symmetries~\cite{Ciliberto1984prl,Meron1987pra}.

The interplay between two pattern-forming instabilities that share the same spectral properties, however, has not been studied. The reason may be the fairly simple pattern forming systems that have been considered in model studies, which do not capture more than one  mechanism for any instability, or to the focus on a single field, rather than on the relations between two independent fields, in empirical studies.

In this paper we use dryland vegetation as a case model for studying the behavior near a codimension-2 point where two instabilities, sharing the same spectral properties but driven by distinct mechanisms, coincide. We focus on the infiltration and uptake-diffusion feedbacks in 1d, which both lead to finite-wavenumber stationary instabilities but result in distinct patterns, in-phase and anti-phase, respectively. Surprisingly, we find that the instability at this point is a codimension-1 bifurcation that leads to the growth of a \emph{single} mode that is neither in-phase nor anti-phase. Nevertheless, the instability does contain information about the two distinct modes - the band of stable periodic solutions that appear beyond the instability point describes a family of stationary periodic patterns ranging \emph{continuously} from in-phase to anti-phase patterns. This behavior is unlike codimension-2 bifurcations, which only show the distinct modes and combinations thereof (e.g. mixed-mode patterns) and 
exclude the range of patterns in between.

Although we address a specific physical context we believe that the main conclusions are general and relevant to other pattern-forming systems too. The vegetation context is particularly appealing because the mechanisms that induce the instabilities are well understood~\cite{Rietkerk2002an,Gilad2004prl,Gilad2007jtb,vanderStelt2012nonl_sci} and the ecological implications are significant as they bear on pattern diversity, at least on ecological time scales~\cite{fn3}, which is a driver of biodiversity~\cite{Shachak2005inbook}.

We study a simplified dimensionless version of the vegetation model introduced in Ref. ~\cite{Gilad2004prl,Gilad2007jtb}, which still captures the infiltration and the uptake-diffusion feedbacks and therefore nonuniform stationary instabilities to in-phase and anti-phase patterns. The model consists of three fields, the areal density of the above ground vegetation biomass, $b(x,t)$, the areal density of soil water, $w(x,t)$, and the areal density of the overland or surface water, $h(x,t)$, which for a flat terrain satisfy the equations:
\begin{subequations}\label{eq:NDM}
  \begin{align}
    \label{eq:dbdt}
      b_t & = g_b b(1-b/\kappa) - b + \nabla^2 b\,, \\
    \label{eq:dwdt}
      w_t & = \mathcal{I}h - \nu w(1-rb/\kappa) -g_w w + \delta_w \nabla^2 w\,, \\
    \label{eq:dhdt}
      h_t & = p - \mathcal{I} h + \delta_h \nabla^2 (h^2)\,,
  \end{align}
\end{subequations}
where $g_{b}= \nu w (1+\eta b)^2$, $g_w=\nu b (1+\eta b)^2$ are the rates of biomass growth and water uptake, respectively, and
\begin{align}
 \mathcal{I}=\alpha \frac{b+q\left(1-\phi\right)}{b+q}\,,
 \end{align}
is the infiltration rate.
The infiltration feedback is captured by the biomass-dependent infiltration rate
$\mathcal{I}$ and the transport term, $\delta_h\nabla^2 (h^2)=-\nabla\cdot
\mathbf{J}, ~~\mathbf{J}=-2\delta_h h\nabla h$, in the equation for $h$, which
describes overland flow along surface-water gradients induced by the high
infiltration rates in vegetation patches. The strength of the infiltration
feedback is controlled by the infiltration contrast parameter $\phi\in[0,1]$ and
the water transport coefficient $\delta_h$. The uptake-diffusion feedback is
captured by the biomass-dependent water-uptake term $-g_w w$, which accounts for
soil-water depletion in patches of growing vegetation, and the diffusion term
$\delta_w \nabla^2 w$, which accounts for soil-water diffusion towards these
patches. The strength of this feedback is controlled by the parameter $\eta$, a
measure for the root-to-shoot ratio, and by the soil-water diffusivity
$\delta_w$. Other model parameters include the precipitation rate $p$, the
evaporation rate of soil water, $\nu$, reduction of evaporation by shading, $r$,
and ``biomass diffusion'' constant, $\delta_b$, which represents clonal growth or
short-range seed dispersal.
We refer the reader to the Supplementary Material
\cite{SM} for the derivation of the simplified model, Eqs.
\eqref{eq:NDM}, and for the relations between the dimensionless quantities appearing in the model
and their dimensional counterparts. More details about the original model can be found in Refs. ~\cite{Gilad2007jtb,Meron2011mmnp}.

Equations \eqref{eq:NDM} have a nonzero stationary uniform solution that represents uniform vegetation. Both the infiltration feedback and the uptake-diffusing feedback can destabilize the uniform vegetation solution. This has been shown using models that capture only one of the two feedbacks~\cite{Rietkerk2002an,vanderStelt2012nonl_sci} and is also shown in Fig. \ref{fig:profiles2} using the model equations \eqref{eq:NDM} that capture both feedbacks. Shown in the figure are results of a linear stability analysis, carried out once when the infiltration feedback is switched off by setting the infiltration
contrast to zero, $\phi=0$ (panels (a), (b) and (c)), and once  when the uptake-diffusion feedback is switched off by setting $\eta=0$ (panels (d), (e) and (f)). In both cases the destabilization of uniform vegetation occurs through a stationary nonuniform instability, characterized by a real-valued eigenvalue attaining a maximal value at a finite wavenumber as the growth-rate curves shown in panels (a) and (d) indicate, but the periodic patterns that appear are different. When the instability is driven by the uptake-diffusion feedback the soil-water content in a patch of denser biomass decreases and the biomass and soil-water distributions are \emph{anti-phase} (panel (b)). When the instability is driven by the infiltration feedback the soil-water content in a patch of denser biomass increases because of the increased infiltration rate, and the distributions are \emph{in-phase} (panel (e)). Panels (c) and (f) show the neutral stability curves for the uptake-diffusion and the infiltration feedbacks 
respectively.

\begin{figure*}[!ht]
  \centering
  \includegraphics[width=16cm]{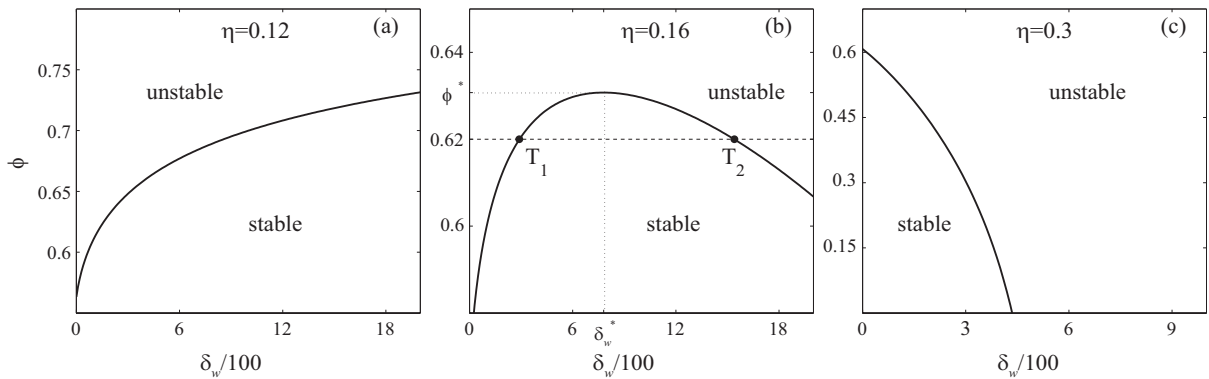}
  \caption{Threshold lines for the nonuniform stationary instability of the uniform state at different $\eta$ values representing (a) the dominance of the infiltration feedback (small value, $\eta=0.12$), (c) the dominance of the uptake-diffusion feedback (large value, $\eta=0.3$), and (b) comparable influence of the two feedbacks (intermediate value, $\eta=0.16$). The threshold lines separate stability and instability domains of the uniform state as denoted. 
  The maximum point of the threshold line, $(\delta_w^*,\phi^*)$, corresponds to a codimension-2 point at which the instability that is driven by the infiltration feedback at $T_1$ merges with that driven by the uptake-diffusion feedback at $T_2$ into a single codimension-1 bifurcation.
  Parameters: $\delta_h=10^4$.}
  \label{fig:dw_vs_phi}
\end{figure*}

In general the two feedbacks act in concert and may affect one another. We studied the interplay between the two feedbacks by exploring the instability threshold of the uniform state in a plane spanned by the parameters $\delta_w$ and $\phi$ that control the uptake-diffusion and infiltration feedbacks, respectively. Figure \ref{fig:dw_vs_phi} shows the instability thresholds for different values of $\eta$. We recall that the parameter $\eta$ controls the strength of the uptake-diffusion feedback (along with $\delta_w$) and is used here to change the relative strength of the two feedbacks.
Panel (a) in Fig. \ref{fig:dw_vs_phi} shows the instability threshold for a relatively low $\eta$ value for which the instability is driven by the infiltration feedback.
As the monotonously increasing threshold line indicates, the alternative uptake-diffusion feedback counteracts the infiltration feedback by inducing soil-water diffusion from water-rich vegetation patches to their dryer neighborhoods, and shifts the instability threshold to higher infiltration contrasts $\phi$.
Panel (c) shows the instability threshold for a relatively high $\eta$ value for
which the instability is driven by the uptake-diffusion feedback. In this case
the threshold line is monotonously decreasing, indicating that the alternative
infiltration feedback promotes the instability by lowering down its threshold.
This behavior can be understood as follows. A higher infiltration contrast
results in the interception of more runoff in denser vegetation patches, which
increases vegetation growth and soil-water uptake and therefore facilitates the
instability by the uptake-diffusion feedback.

At intermediate $\eta$ values both feedbacks are equally important and the
instability threshold line is no longer monotonous as panel (b) in Fig.
\ref{fig:dw_vs_phi} shows.
High values of $\delta_w$ lead to an instability of the uniform vegetation state by the uptake-diffusion feedback and to the formation of an anti-phase periodic pattern (point $T_2$ in the diagram).
Low values of $\delta_w$ lead to an instability of the uniform state by the infiltration feedback and to the formation of an in-phase periodic pattern (point $T_1$ in the diagram).
The instability at $T_1$ occurs despite the fact that the parameters that control it, $\phi$ and $\delta_h$, are held constant. This is because of the counter effect that the uptake-diffusion feedback has on the infiltration feedback.

\begin{figure*}[!ht]
  \centering
  \includegraphics[width=16cm]{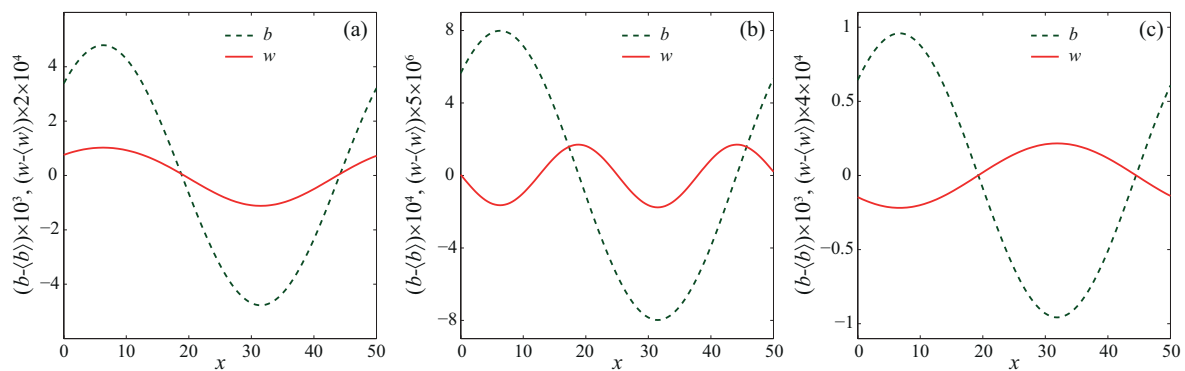}
  \caption{The stationary periodic patterns that develop beyond the instabilities of the uniform state. Panels (a) and (c) show in-phase and anti-phase patterns obtained by instabilities driven by the infiltration and uptake-diffusion feedbacks, respectively. Panel (b) shows the pattern that results beyond the merging point of the two instabilities. The pattern is neither in-phase nor anti-phase showing maximal soil-water content at the two rims of a biomass patch. }
  \label{fig:PhaseChange}
\end{figure*}

\begin{figure*}[!ht]
   \centering
   \includegraphics[width=0.9\linewidth]{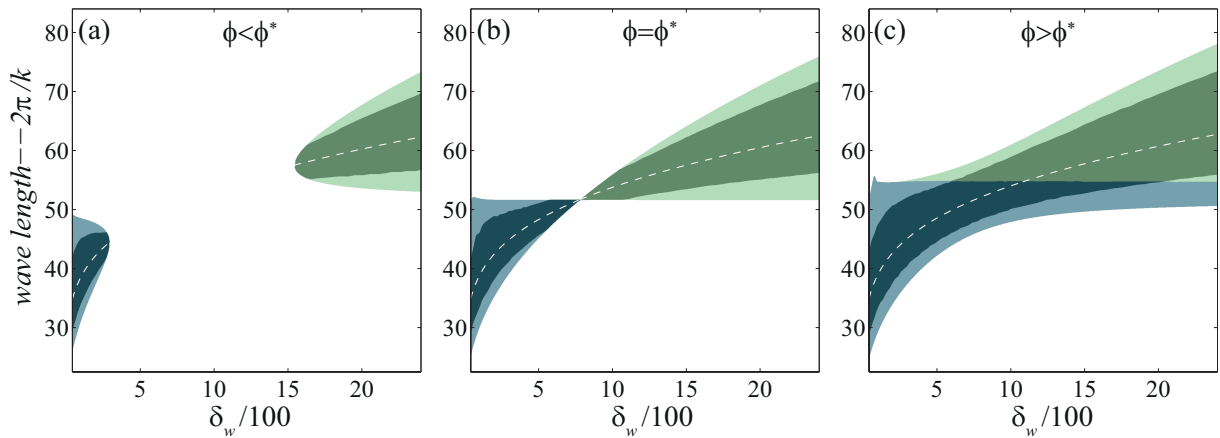}
   \caption{Busse balloons of periodic solutions below ($\phi<\phi^*$), at ($\phi=\phi^*$) and above ($\phi>\phi^*$) the codimension-2 point (see Fig. \ref{fig:dw_vs_phi}(b) for the definition of $\phi^*$). The blue (green) shade represents the existence range of in-phase (anti-phase) solutions. The darker shades denote stable solutions. The dashed line denotes the wavelength that corresponds to the maximal growth rate as  calculated by the linear stability analysis. Above the codimension-2 point ($\phi>\phi^*$) a band of stationary periodic patterns, ranging continuously from in-phase to anti-phase patterns, exists. The stability of the solutions was calculated using numerical linear stability analysis, where a domain length of $10L$ was taken (where $L=2\pi/k$ is the wavelength of the solution).}
   \label{fig:ExiStaBalloon}
 \end{figure*}

Interestingly, we find that there is a particular point $(\delta_w^*, \phi^*)$ in the ($\delta_w,\phi$) plane at which the instabilities at $T_1$ and at $T_2$ coincide, but, contrary to what one might expect, this is not a codimension-2 bifurcation; the instability at  $(\delta_w^*, \phi^*)$ is a codimension-1 bifurcation characterized by the growth of a single mode. As Fig. \ref{fig:PhaseChange} shows the growth of this mode results in a periodic pattern that is neither in-phase nor anti-phase. We call this pattern a ``rim pattern'' because the soil-water distribution has maxima at the two rims of each biomass hump or patch.
To better understand the interplay between the two pattern-forming feedbacks away and in the vicinity of the point $(\delta_w^*,\phi^*)$, we used a numerical continuation method to calculate the existence boundaries of periodic solutions with different wavenumbers $k$, and numerical stability analysis (in 1d) to evaluate their stability thresholds. Results of this analysis for three different values of $\phi$ are shown in Fig. \ref{fig:ExiStaBalloon} in the form of ``Busse balloons'', i.e. as graphs of solution wavelength vs. the control parameter $\delta_w$~\cite{vanderStelt2012nonl_sci}. The blue (green) shades denote existence ranges of solutions representing \textit{in-phase} (\textit{anti-phase}) patterns, and the dark shades denote stability ranges of these solutions. When $\phi<\phi^*$ (panel (a)) the uniform state is stable for intermediate $\delta_w$ values and loses stability to anti-phase (in-phase) patterns as $\delta_w$ is increased (decreased) past a threshold value. The Busse balloons 
associated with the two instabilities are separate, implying the existence of either in-phase patterns or anti-phase patterns depending on the value of $\delta_w$. When $\phi=\phi^*$ (panel (b)) the two Busse balloons touch one another at the codimension-2 point, and when $\phi>\phi^*$ (panel (c)) they overlap. As the dark shades indicate, there exists a \emph{continuous} band of stable periodic patterns ranging from in-phase patterns at low wavelengths to anti-phase patterns at high wavelengths that includes the rim pattern that is neither in-phase nor anti-phase (dashed line). This pattern diversity is higher than the diversity of patterns that would have resulted from a codimension-2 bifurcation (simultaneous growth of distinct in-phase and anti-phase modes) because it includes all intermediate patterns.

The predicted multiplicity of stable biomass-water patterns ranging continuously from in-phase to anti-phase patterns bears on the biodiversity of water-limited ecosystems. Water limited landscapes often consist of woody and herbaceous vegetation (e.g. shrubs and annuals). The woody species generally form spatial patterns of biomass and soil water to which the herbaceous community responds. The continuous range of the soil-water distributions formed by the multiplicity of stable woody patterns provides a wide variety of complementary habitats for herbaceous species. These habitats consist of water-rich areas and divide into three major classes: Open areas between woody patches (anti-phase pattern), woody-patch areas (in-phase patterns), and rims of woody patches (rim patterns). Landscapes in which all pattern types appear can accommodate high species diversity because they provide habitats that make various compromises in terms of exposure to light, nutrients, grazing, etc.

An interesting mathematical question that can shed light on the generality of the results is whether the co-occurrence of two or more instabilities that share the same spectral properties (eigenvalues of the linear problem) necessarily implies a codimension-1 bifurcation that involves the growth of a single mode.
\begin{acknowledgements}
The research leading to these results has received funding from the European Union Seventh
Framework Programme (FP7/2007-2013) under grant number 293825, and from the Israel Science Foundation, under Grant no. 861/09.
\end{acknowledgements}


\begin{thebibliography}{14}
\bibitem[{fn1()}]{fn1}
 \bibinfo{title}{A ``codimension-2 point'' is a point in the plane of two control parameters that represents the threshold values of the instabilities they control}.
\bibitem[{fn2()}]{fn2}
 \bibinfo{title}{We distinguish here between ``codimension-2 points'' and ``codimension-2 bifurcations''. The former term merely denotes a point in a plane spanned by two control parameters, whereas the latter term refers to a codimension-2 point where two instabilities that give rise to distinct modes coincide}.
\bibitem[{fn3()}]{fn3}
 \bibinfo{title}{Ecosystems are generally subjected to disturbances, which limit the time scales of undisturbed ecosystem dynamics and often make them significantly shorter than the time scales needed to approach asymptotic states}.
\end{thebibliography}
\end{document}